\documentclass[3pt,onecolumn]{article}
\usepackage{graphicx}
\usepackage{epstopdf}

\usepackage{color}

\newcommand{\ben}{\begin{enumerate}}
\newcommand{\een}{\end{enumerate}}

\newcommand{\be}{\begin{equation}}
\newcommand{\ee}{\end{equation}}
\newcommand{\bea}{\begin{eqnarray}}
\newcommand{\eea}{\end{eqnarray}}

\begin{document}

\vspace{0.1cm}
\thispagestyle{empty}


\begin{center}

{\Large\bf The size of the nucleosome}\\[13mm]

{\it Jakob Bohr and Kasper Olsen}\\
{\it DTU Nanotech, Building 345 East, \O rsteds Plads}\\
{\it Technical University of Denmark, 2800 Kongens Lyngby, Denmark}\\
jabo@nanotech.dtu.dk \& kasol@nanotech.dtu.dk\\[6mm]
\end{center}

\begin{abstract}
The structural origin of the size of the 11 nm nucleosomal disc is 
addressed. On the nanometer length-scale the organization of DNA as chromatin in the chromosomes involves a coiling of DNA around the histone core of the nucleosome. 
We suggest that the size of the nucleosome core particle is dictated by
the fulfillment of two criteria:
One is optimizing the volume fraction of the DNA double helix; this requirement for close-packing has its root in optimizing atomic and molecular interactions. The other criterion being that of having a zero strain-twist coupling; being a zero-twist structure is a necessity when allowing for transient tensile stresses during the reorganization of DNA, e.g., during the reposition, or sliding, of a nucleosome along the DNA double helix. The mathematical model we apply is based on a tubular description of double helices assuming hard walls. When the base-pairs of the linker-DNA is included the estimate of the size of an ideal nucleosome is in close agreement with the experimental numbers. Interestingly, the size of the nucleosome is shown to be a consequence of intrinsic properties of the DNA double helix.

\end{abstract}\vspace{0cm}

\addtocounter{section}{1}

\section*{Introduction}

The packing of DNA and histones into nucleosomes is the first level of compaction of the deoxyribonucleic acid of the chromosomes. The nucleosomes form the fundamental repeating building structure of chromatin, each incorporating a stretch of about two-hundred consecutive base pairs \cite{kornberg1974}. A nucleosome consist of the core nucleosome particle with 146/147 bp and a linker-DNA of variable length, typically with about 50 bp \cite{widom1992}. Crystallographic studies have revealed the detailed geometry of the 
nucleosome core particle \cite{finch1977,luger1997,richmond2003}.  An octameric disc is formed of the alkaline proteins H2A, H2B, H3 and H4; on this disc the double stranded helical DNA is coiled $\sim 1.7$ times in a nearly flat left-handed superhelix. Cells in the human body contain $2 \times 23 $ chromosomes with perhaps as many as $4 \times 10^7$ nucleosomes in total, and most remarkable these four histones are nearly conserved across all eukaryotic cells. Even in cases with significant sequence differences they remain structurally exceptionally similar \cite{white2001}. A typical yeast genome contains about $6 \times 10^4$  nucleosomes \cite{rando2010}. 

Structurally, the nucleosomes support an efficient compaction of DNA \cite{zinchenko2006}. Electrostatic interactions between DNA and the histones contribute to the stability of the nucleosome \cite{west2010,fenley2010}, as well as of any formed superstructures such as the 30 nm filament \cite{widom1986,horowitz1994,bednar1998}. During the last decade, or two, it has become clear that the functional roles of chromatin/nucleosomes/histones are much richer than being limited to that of compaction \cite{wyrick1999,berger2007,shukla2009}. Post-translational biochemical processes \cite{ausio1986,strahl2000,turner2000,sekinger2005,shilatifard2006}, e.g. methylation, ubiquitination, and acetylation, as well as nucleosome-positioning proteins \cite{strauss1984} can have profound and decisive impact on regulating phenomena such as transcription, recombination, and repair.  Biochemical processes may also control phenomena such the sliding of the nucleosomes along the DNA \cite{schiessel2001,zlatanova2003} and of possible histone eviction \cite{schwabish2006}. It was early recognized that the DNA of the nucleosome has a modified number of base pairs per full rotation \cite{levitt1978,klug1981,hayes1990}. The average number is about 10.1 bp/$2\pi$  for nucleosomal DNA in comparison with 10.5 bp/$2\pi$ for B-DNA.

The linking number of a set of two closed space curves is a conserved integer. White's theorem \cite{white1969} equate the topological linking number with the sum of two geometrical measures -- the writhing and the total twist. For a review of the relevance of the linking number for biomolecules and of knot theory see ref. \cite{skjeltorp1996}. In practice the linking is also conserved for a local stretch of a curve when the full curve extends too far for it to globally reorganize. An example of such a local event is the eviction of the histones associated with the dissolution of one nucleotide \cite{schwabish2006}. 
Models of the chromatin fiber have incorporated simplified interactions between nucleosome core particles \cite{besker2005}. Recently, 
the understanding of linking as well as of steric interactions have been further embraced in proposed detailed models for the chromatin fiber \cite{wong2007}. Furthermore, models of the chromatin fiber has incorporated detailed molecular force fields \cite{kepper2008}.

In this paper we address the size of the nucleosome with its structurally highly conserved core. It is interesting to inquire into why the nucleosome core particles have the size they have, and to know the answers to related questions. For example, would it at all be possible for the nucleosomes to be twice the size?  We propose that the DNA of the nucleosomes must obey certain criteria and show that the combination of two criteria leads to a unique requirement for the size of an ideal nucleosome. This can explain why the size of the nucleosomal disc is highly conserved throughout the eukaryotes.

\section*{Model}

\noindent 

In the following we define two criteria, one for the twisting behavior and one for the packing properties and suggest that nucleosomal DNA obeys both of the criteria. We then model an ideal nucleosome as one that obeys the two criteria simultaneously. To find the size of an ideal nucleosome it is necessary to understand how the two criteria changes with the superhelical curvature of the DNA double helix, i.e. with bending of the double helix. To demonstrate this behavior we calculate the results for tubular double helices. This simplification has worked well in the past for generic $\alpha$-helices, and for A- and B-DNA where accurate theoretical estimates of the pitch angles of experimental structures are obtained \cite{olsen2009}.

\noindent Here the criteria are discussed in connection with their application with long polymers:

\noindent {\bf The Zero-Twist Criterion.} {\it
When long structures are under longitudinal strain they tend to twist. Certain structures have a vanishing strain-twist coupling, we denote such structures as {\it zero-twist} structures. 
Fascinatingly, even chiral structures can be zero-twist structures \cite{olsenZT}. The zero-twist criterion simply requires structures to be zero-twist structures, as this prevents a rotational catastrophe for long structures. Recently, we have shown that $N$-helices with a unique pitch are zero-twist structures \cite{olsenZT} and  have suggested that it is advantageous for the triple helix of collagen to obey the zero-twist criterion \cite{Bohrcollagen}.}

\noindent {\bf Close-Packing Criterion.} {\it Atomic and molecular structures tend to use space efficiently, e.g. a good fraction of the elements of the periodic system form close-packed structures, or nearly close-packed structures. Recently, we have shown that not all helices use space equally efficient \cite{olsen2009}; the helices that optimize the volume fraction are denoted close-packed helices. The $\alpha$-helices, A-DNA, and B-DNA have all been shown to obey the close-packed criterion \cite{olsen2009}. In short, it is difficult to increase the effective volume of an atomic or a molecular structure even by a relatively tiny fraction. This is the origin of the close-packing criterion.}

Being zero-twist and having a conserved linking number both relate to the twisting of a polymer chain. However, the two features are distinct from each other. Having a conserved linking number says that the oriented space curve of the polymer backbone shall preserve its linking, unless enzymatic cleaving of the sugar-phosphate backbone of DNA is active. 
The difference between zero-twist and conserved linking number can be understood when one considers a transformation from molecular {\it path-I} to {\it path-II}. Such an actual transformation will typically involve a transient longitudinal straining of the molecular structure. This is where the zero-twist criterion comes into play: It prevents a transient catastrophic twisting of long polymers. An example which illustrates this is the sliding (or repositioning)  of a nucleosome along the double stranded DNA polymer. Such a repositioning of the nucleosome trivially obeys having a conserved linking number. This is not so for the zero-twist criterion, generally a strained helical structure will twist, i.e. long-ranged and accumulative modifications to the helix will result from an induced longitudinal strain. Certain unique structures have zero strain-twist coupling, these are the zero-twist structures that can be longitudinally strained without causing long-ranged propagated modifications in twist.

\section*{Results}

First we consider the case of absence of superhelical curvature, i.e. a straight section of a double helix. Figure 1A shows the calculated volume fraction, $f_V$, as function of the dimensionless value of the double helical pitch, $H/D$. Here, $H$ is the pitch of the double helix and $D$ the diameter of one of the two strands of the double helix. For DNA the strand diameter is $D \approx 11$ \AA~\cite{olsen2009}. 
The maximum value of the volume fraction, see Figure 1A, is obtained for $H/D$=2.399, this double helix obeys the close-packing criterion. Likewise, the twist, $\Theta$, of a straight section of a double helix can be calculated. It is more useful to calculate the twist normalized by the dimensionless length of the strands, $D\Theta /L$. Figure 1B depicts $D \Theta /L$ as a function of $H/D$.  The twist is the incremental rotation of the double helix as one progresses along one of the two strands. The maximum appears at $H/D$=2.697, this double helix obeys the zero-twist criterion.

\begin{figure}[h]
\begin{center}
\includegraphics[width=7cm]{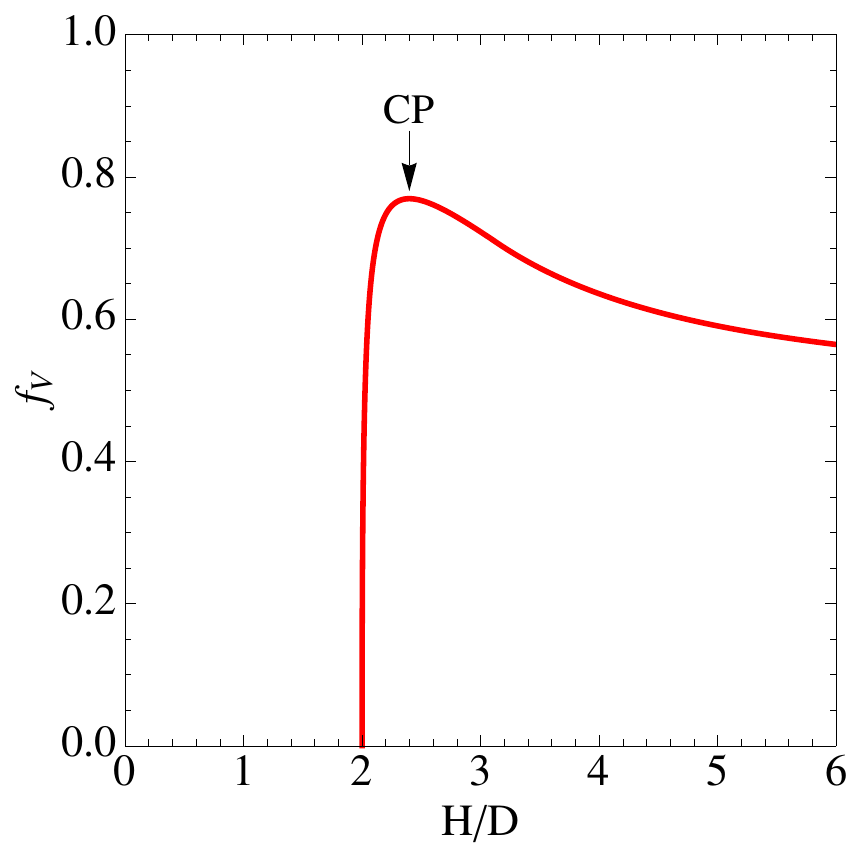}\includegraphics[width=7cm]{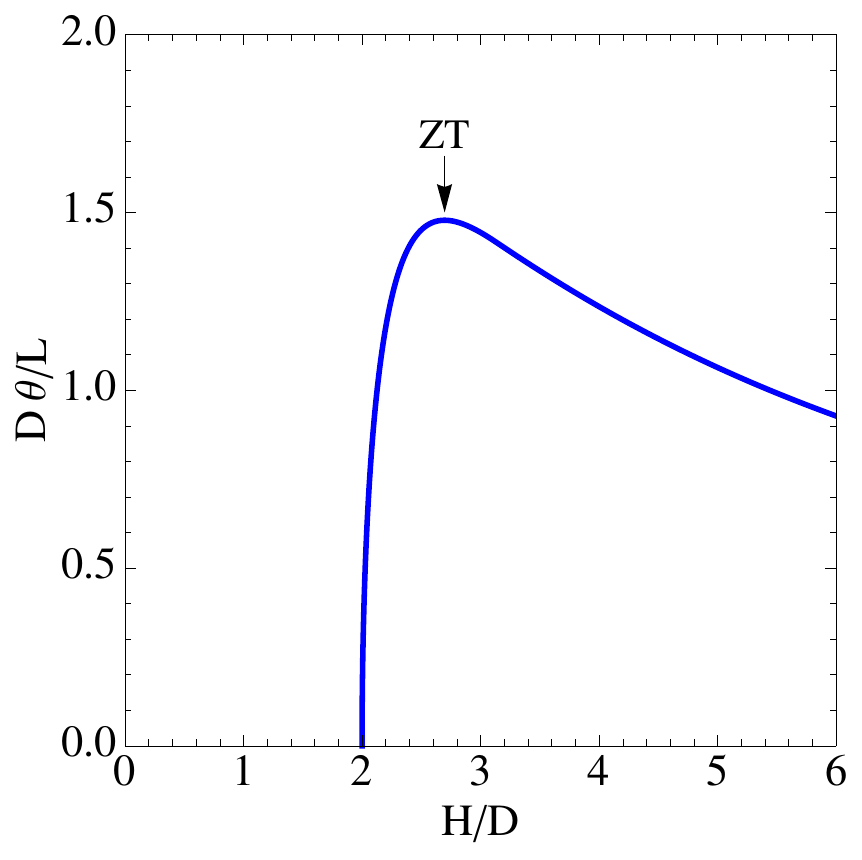}
\caption{\it (A) The volume fraction $f_V$ for  a symmetric tubular double helix as a function of the dimensionless pitch, $H/D$, where $H$ is the pitch (the length of a repetitive segment of the double helix), and $D$ the diameter of one of the tubular strands. The maximum value of the volume fraction is $0.769$ for $H/D=2.399$. This is the close-packed double helix with a pitch angle of $32.5^\circ$ \cite{olsen2009}. (B) The rate of change, $D \Theta /L$, in the twist of a tubular double helix per unit increment of the dimensionless length of the strands, $L/D$, depicted as a function of the dimensionless pitch, $H/D$, of the helix. The maximum is at $H/D=2.697$.}
\end{center}
\end{figure}


Therefore it is not possible for a straight double helix to simultaneously obey the zero-twist criterion (response to strain) and the close-packing criterion (molecular forces). A compaction of DNA that involves long straight sections of DNA would therefore exhibit a rotational catastrophe under tensile stress. However, the nucleosomes bend the DNA and thus introduce a curvature to the double helix as the double helix is coiled around the histone proteins. As curvature is introduced, the steric interactions between the two individual strands of the double helix changes. This means that the curves depicted in Figure 1A and Figure 1B needs to be recalculated for each value of the curvature of the center line of the double-stranded DNA molecule. Hence, the pitch of the helix for which the zero-twist criterion is fulfilled changes with the amount of introduced curvature, correspondingly so for the close-packing criterion. Figure 2 shows the pitch, $H/D$,  of the zero-twist and close-packed double helices as a function of $D/R$. The radius of curvature, $R$, is the radius of the central line of the double helix of nucleosomal DNA. A straight segment of DNA has $R=\infty$. For both the zero-twist helix and the close-packed helix the helical pitch is increased when bending is enforced, though more rapidly so for the close-packed helix. The result is that for a curvature of $D/R$=0.391 the two criteria can be simultaneously obeyed. We suggest that this is the defining science behind the size of the nucleosome. We therefore estimate the size of such an ideal nucleosome to be $2R = 56$~\AA. 

\begin{figure}[h]
\begin{center}
\includegraphics[width=8cm]{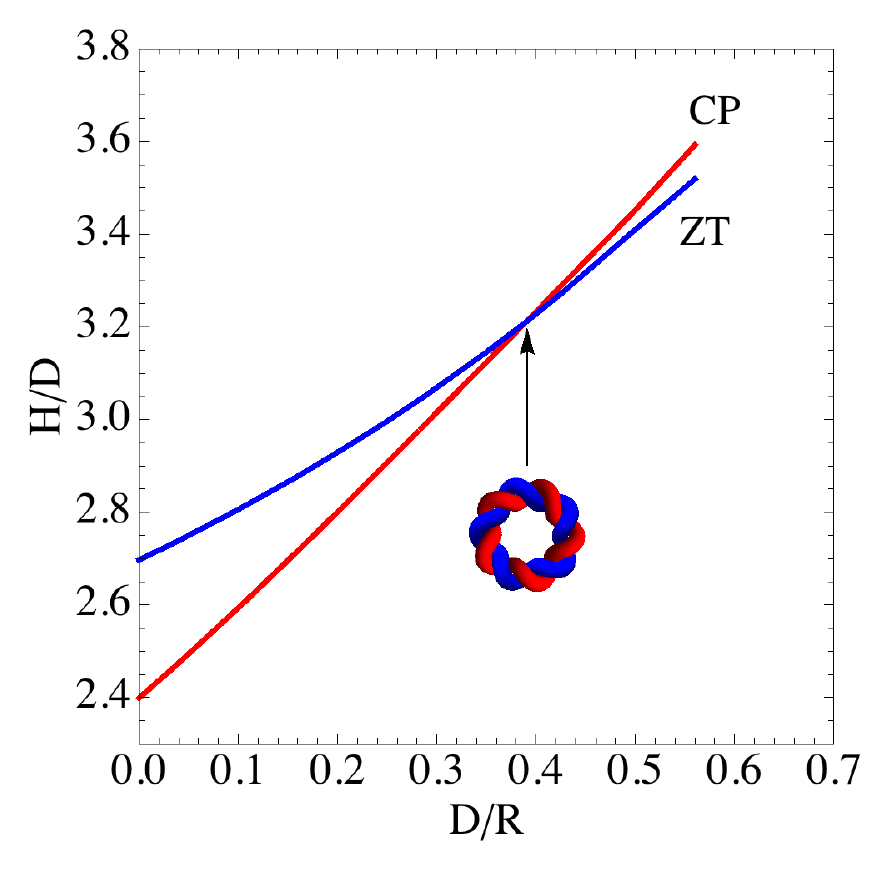}
\caption{\it The blue curve shows the value of the helical pitch that obeys the zero-twist criterion as a function of the dimensionless curvature, $D/R$, and the red curve shows the value of the helical pitch for the curved helices that obey the close-packed criterion as a function of  curvature, $D/R$. As can be seen the two curves meets at a value of $D/R=0.391$, with $D \approx11$ {\rm \AA}~this corresponds to $2R=56$ {\rm \AA}. The torus double helix shown has $D/R=0.391$ and is therefore both close-packed and zero-twist.}
\end{center}
\end{figure}

\begin{figure}[h]
\begin{center}
\includegraphics[width=7cm]{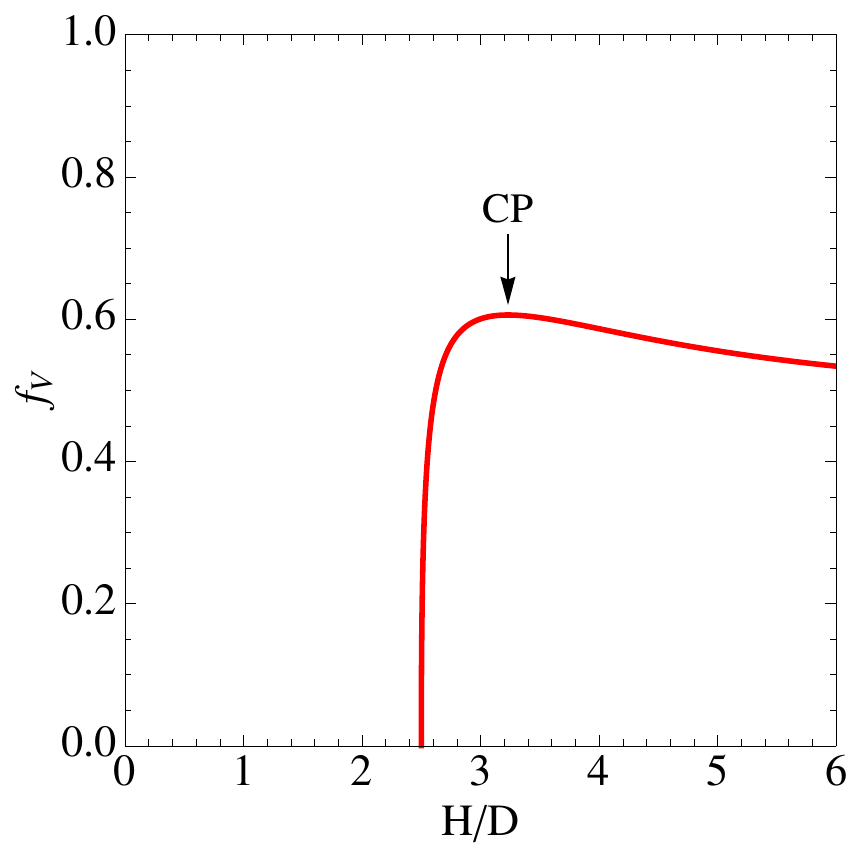}\includegraphics[width=7cm]{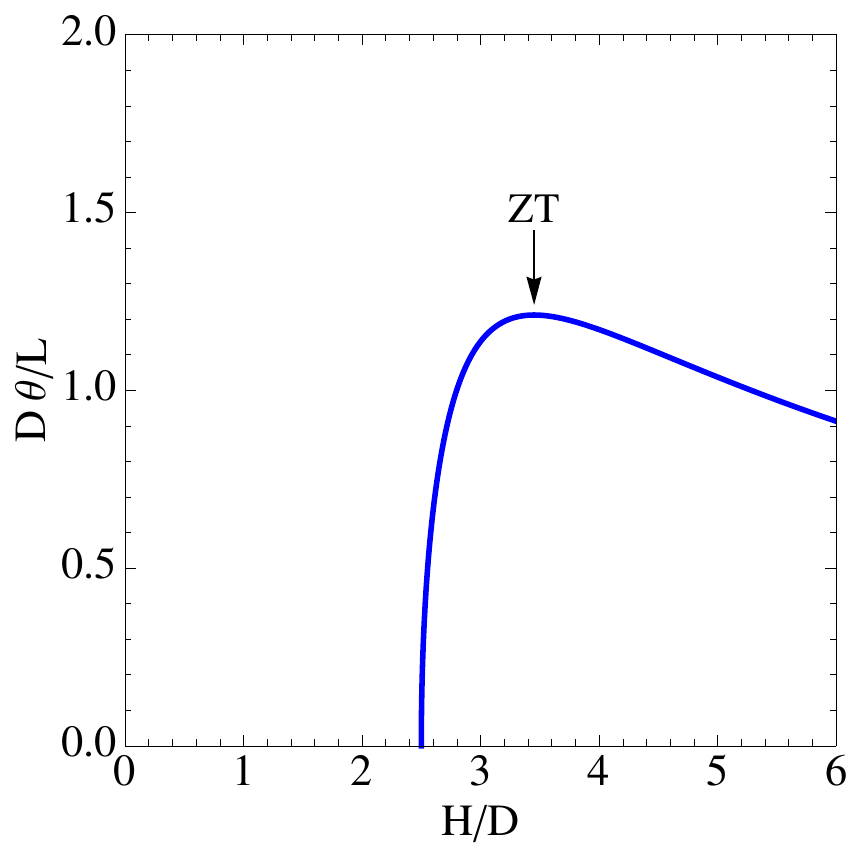}
\caption{\it (A) The volume fraction $f_V$ for  an asymmetric (minor groove $144^\circ$) tubular double helix as a function of the dimensionless pitch, $H/D$, where $H$ is the pitch (the length of a repetitive segment of the double helix), and $D$ the diameter of one of the tubular strands. The maximum value of the volume fraction appears for $H/D=3.232$. This is the close-packed double helix with a pitch angle of 38.1$^\circ$. In ref. \cite{olsen2009} a value $142^\circ$ was used for the minor groove and therefore the obtained pitch angle was $38.5^\circ$ instead of $38.1^\circ$. (B) The rate of change, $D \Theta /L$, in the twist of an asymmetric (minor groove 144$^\circ$) tubular double helix per unit increment of the dimensionless length of the strands, $L/D$, depicted as a function of the dimensionless pitch, $H/D$, of the helix. The maximum value is at $H/D=3.452$.}
\end{center}
\end{figure}

The above results only holds for symmetric double helices. B-DNA is asymmetric with 144$^\circ$ between the strands towards the minor groove, and 216$^\circ$ for the major groove \cite{klug1981}. The presented calculation for symmetric double helices therefore needs to be modified with steric restrictions for the asymmetric double helix. Figure 3A shows the volume fraction, $f_V$, for an asymmetric helix and Figure 3B shows the rate of increment in twist, $D\Theta /L$, for an asymmetric double helix in the absence of curvature.
Figure 4 shows the dimensionless pitch, $H/D$, of the asymmetric B-DNA like double helix as a function of introduced curvature, $D/R$. The results is now that for a curvature of $D/R=0.141$ the two criteria can be simultaneously obeyed. This leads to an estimate for $2R=156$~\AA . 


\begin{figure}[h]
\begin{center}
\includegraphics[width=8cm]{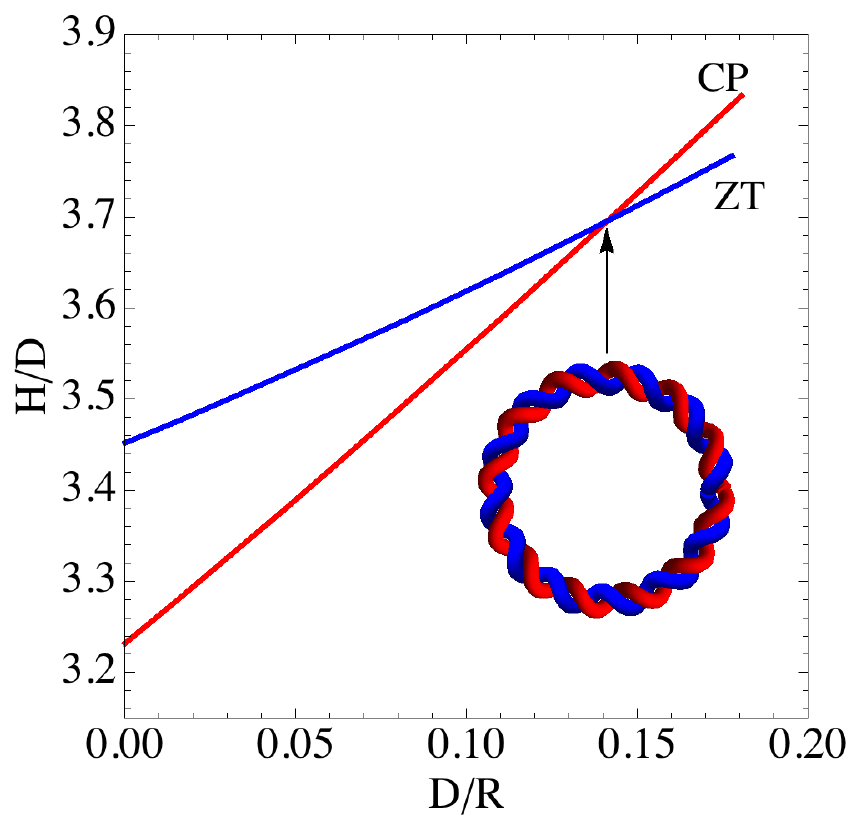}
\caption{\it The blue curve shows the value of the helical pitch of helices that obey the zero-twist criterion as a function of the dimensionless curvature, $D/R$, and the red curve shows the value of the helical pitch for the curved helices that obey the close-packing criterion as a function of  curvature, $D/R$. As can be seen the two curves meets at a value of $D/R=0.141$, with $D=11$ {\rm \AA}~this corresponds to $2R = 156$ {\rm \AA}. The asymmetric double helix also has $D/R=0.141$ and is therefore both close-packed and zero-twist.}
\end{center}
\end{figure}

We now take into account that the nucleosome core particle comes with an additional stretch of B-DNA, often denoted the linker-DNA. If one makes the requirement that the nucleosome core particle together with the linker-DNA behaves as a zero-twist structure then the nucleosome core particle needs to properly compensate for the behavior of the linker-DNA. I.e. the nucleosome core particle needs to have a further reduced radius. B-DNA has a incremental twist of $D^2 d\frac{\Theta}{L} / dH= 0.135$ at the close-packed structure. If we assume that the linker-DNA is about $50$ bp then the DNA of the nucleosome core particle needs to have an incremental twist of $D^2 d \frac{\Theta}{L} / dH = - (50/147) 0.135 = -0.046$. This requirement leads to $D/R=0.221$ and a diameter of $2R=100$~\AA . For this particular curvature of DNA the twisting, and counter-twisting, of the linker-DNA and of the nucleosomal core particle DNA precisely cancel. Table 1 summarizes the scale invariant parameters for the three presented nucleosomal models. Three parameters fully describe a structure, the fourth parameter is given for convenience. Models of chromatin often incorporate a relatively large range for the length of the linker-DNA (30-90 bp) \cite{wong2007,kepper2008}. If we use the less common value 70 bp we find $2R=83$~\AA~and a disc diameter of 11 nm. 

\begin{table}
\caption{\it Calculated values for the scale invariant variables describing the bent symmetric double helices (A-DNA) which has both close-packed (CP) and zero-twist (ZT) properties. Same for the bent asymmetric double helix (B-DNA), here the phase of the minor groove is set at $144^\circ$ \cite{klug1981}. Lastly, is listed the values for the curved asymmetric double helices (B-DNA) which compensates the twisting such that it precisely cancels a 50 bp stretch of linker-DNA.}
\label{tab:1}       
\begin{tabular}{llllll}
\hline\noalign{\smallskip}
Model & Description & $D/R$ & $H/D$ & $2a/D$ & $L_{2\pi}/D$\\
\noalign{\smallskip}\hline\hline\noalign{\smallskip}
Model 1 & CP \& ZT curved A-DNA &  0.391 & 3.212 & 1.029 & 4.569\\

Model 2 & CP \& ZT curved B-DNA & 0.141 & 3.696 & 1.237& 5.364\\

Model 3 & CP \& compensating B-DNA & 0.221& 3.976& 1.203& 5.493\\

\noalign{\smallskip}\hline
\end{tabular}
\end{table}

\section*{Discussion}

Measuring the distance from the centerline of the DNA helix of the nucleosome core particle PDB entry 3LZ0 \cite{vasudevan2010} we find $2R_{PDB} \sim 83$~\AA. Table 2 compares the values of $2R$ from crystallographic studies with the three models discussed above and enlisted in Table 1. The value for the compensating model $2R=100$ \AA~ is in good agreement with the experimental value of 83-84 \AA . The outer diameter of the nucleosomal disc for this model is $2R+2a+D=12$ nm, in fair agreement with the well established size of the nucleosome disc, i.e. 11 nm. 

\begin{table}[b]
\caption{\it Comparison of experimental and calculated values for $2R$ for nucleosomal DNA and for the diameter of the nucleosome disc $2R+2a+D$. Model 1 is for a symmetric double helix (A-DNA). Model 2 is for an asymmetric double helix (B-DNA). Model 3 is for an asymmetric helix including the effect of a 50 bp long linker-DNA. The numbers for $2R$ are given in units of \AA~while the numbers for the disc size, $2R+2a+D$ are in units of nm.}
\label{tab:1}       
\begin{tabular}{lrrrrrr}
\hline\noalign{\smallskip}
 & & 3ZL0 & 1ZLA & Model 1 & Model 2 & Model 3\\
\noalign{\smallskip}\hline\hline\noalign{\smallskip}
Nucleosomal DNA & $2R$ (\AA)   & 83 & 84 & 56 & 156& 100\\
\noalign{\smallskip}\hline\noalign{\smallskip}
Size of the nucleosome & $2R+2a+D$ (nm)& 11  & 11 & 8 &  18& 12\\

\noalign{\smallskip}\hline
\end{tabular}
\end{table}

When one bends a double helix, as is the case when wrapping it around  the histone proteins, curvature is introduced and the steric interactions are modified compared to those in a straight double helical segment. This modifies the number of base pairs per full rotation. Such a change in the number of base pairs per rotation was first observed in theoretical studies using empirical energy functions \cite{levitt1978}. Figure 5 depicts the number of base pairs per full $2\pi$ rotations for the asymmetric tube model as a function of dimensionless curvature $D/R$. As can be seen for the model 3, $D/R = 0.221$, we predict 10.0 bp/$2\pi$ slightly below the experimentally known figure of 10.1 bp/$2\pi$.   The discussion shows the steric interactions within the double helix causes DNA to twist further when it is bent, i.e. to have fewer base pairs per full rotation. This is why the superhelix of the nucleosomal DNA is left-handed, as this handedness helps to compensate for the change in linking. In other words, had the nucleosomal DNA superhelix been right-handed then there would have been a serious linking number discrepancy associated with the evictions of the histones.

\begin{figure}[h]
\begin{center}
\includegraphics[width=8cm]{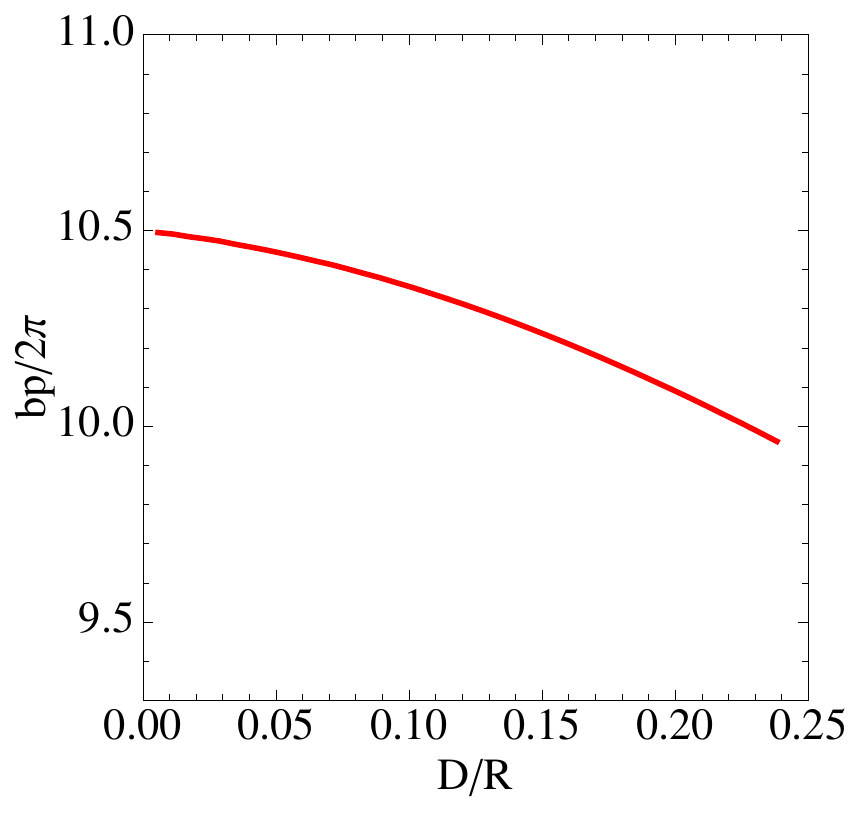}
\caption{\it The estimated number of base pairs per full $2\pi$ rotation for the curved asymmetrical tubular double helix. Normalized such that there is 10.5 bp/$2\pi$ for zero curvature. I.e. for B-DNA $L_{2\pi} /D= 5.238 $ with 10.5 bp/$2 \pi$.}
\end{center}
\end{figure}

By combining two criteria -- to be close-packed and to be a zero-twist structure -- we have come to the suggestion that the DNA of the nucleosome must have a specific curvature. Therefore, neither with a core of histones twice the size, nor with a core half the size, would the nucleosomal construction have worked. This is consistent with the octameric disc of the H2A, H2B, H3 and H4 histones being structurally highly conserved. Furthermore, the use of a tubular model reaches a number for the size which agrees with the scale of known size of the nucleosomes from the PDB data base, and predict the approximately observed change in number of base pairs per full rotation. Considering the geometrical idealization of using a tubular model the agreement is remarkable. It is interesting to note that this means that the two features that determine the size of the nucleosome when working in congruity both are intrinsic features of the double helix of DNA. Hence the presented results could not have been obtained by modeling the double helical DNA as a single tube. There is currently much progress in experimental studies of tension and torsion of chromatin, for a review see \cite{lavelle2010}, and presumably studies on single nucleosome core particles will soon be performed.

\section*{Methods:}

\noindent We utilize a tubular model and assume in all calculations that the helix is formed of two flexible tubular strands with hard walls which are in contact with each other. These contacts define the steric hinderance, and their solutions are found numerically which involves solving transcendental equations. For a detailed description of how we solve the condition of inter-strand contact, see ref. \cite{olsen2009}, wherein are also described how the asymmetry of B-DNA is handled in tubular models. For references to earlier work on how to solve for inter-strand contacts see ref. \cite{pieranski1998,przybyl2001,neukirch2002}.

The twist, $\Theta$ of straight section of a double helix formed from two tubular strands each being $L$ long can be calculated as

\begin{equation}
\Theta = \frac{L}{a} \left(\sqrt{1+(\frac{H}{2\pi a})^2} \, \, \right)^{-1}
\end{equation}

\noindent Here $H$ is the pitch of the double helix and $a$ the radius of the imaginary cylinder which hosts the parametric representation of the strands. In the following equations $D$ is the diameter of one of the two flexible tubes. 

\noindent We define a zero-twist structure as one that obeys,

\begin{equation}
D^2 \frac{d}{dH} (\frac{\Theta}{L}) =0
\end{equation}

\noindent Therefore, a zero-twist structure is one that neither rotates clockwise nor anti-clockwise when under tensile strain.

\noindent The volume fraction of a double helix, $f_V(H/D)$, is easily calculated \cite{olsen2009}. It describes the effectiveness of the use of space calculated relatively to an enclosing cylinder,

\begin{equation}
f_V = \frac{2 V_S}{V_E}
\end{equation}

\noindent where $V_S$ is the volume of a strand being $L$ long, $V_E$ is the volume of the smallest cylinder which circumscribes the double helix section. For a bent double helix, the reference volume becomes a section of a torus instead of a cylinder.

\noindent The criterion for being close-packed is that the volume fraction, $f_V$, is at its maximum. The maximum point can be found by considering,

\begin{equation}
D \frac{d}{dH} (f_V) =0
\end{equation}

\noindent Expressed in terms of the parameters $H$, $a$, and $D$ the volume fraction becomes

\begin{equation}
f_V = 2 \left( \frac{2a}{D} +1\right)^{-2} \left( (\frac{2\pi a}{H})^2 +1\right)^{1/2}
\end{equation}

For the case of a curved section of a double tubular helix the continuous symmetry of the inter-strand contacts point is broken. This means that the inter-strand contact points no longer form helical lines but now appears in discrete sets of points. For the calculation we have consistently used the most restrictive condition of inter-strand contacts. This condition appears when the minor groove coincides with the inside of the equatorial plane of the bent double helix.

\noindent Consistent with this, the twist, $\Theta$, of a bent double helix is

\begin{equation}
\Theta = \frac{L}{a} \left(  \frac{1}{\pi} \int^{\pi}_0 \sqrt{1+(\frac{H}{2\pi a} (1-\frac{a}{R} \cos t))^2} \, \, dt\right)^{-1}
\end{equation}

\noindent where $R$ is the radius of the centerline of the bent double helix. Similarly, the volume fraction of the bent double helix is

\begin{equation}
f_V = 2 \left( \frac{2a}{D} +1\right)^{-2} \frac{1}{\pi} \int^{\pi}_{0}\left( (\frac{2\pi a}{H})^2 +(1-\frac{a}{R} \cos t)^2\right)^{1/2} \, dt
\end{equation}

\hspace{2cm}

\section*{Acknowledgements:} We thank the Villum Foundation for financial support.



\begin{thebibliography}{}

\bibitem{kornberg1974}
Kronberg RD (1974)
{Chromatin structure: a repeating unit of histones and DNA}.
{\it Science} 184:868-871.

\bibitem{widom1992}
Widom J (1992)
{A relationship between the helical twist of DNA and the ordered position of nucleosomes in all eukaryotic cells}.
{\it Proc. Natl. Acad. Sci. USA} 89:1095-1099.

\bibitem{finch1977}
Finch JT, Lutter LC, Rhodes D, Brown RS, Rushton B, Levitt M, Klug A (1977)
{Structure of nucleosome core particles of chromatin}.
{\it Nature} 269:29-36.

\bibitem{luger1997}
Luger K, M\"{a}der AM, Richmond RK, Sargent DF, Richmond TJ (1997)
{Crystal structure of the nucleosome core particle at 2.8 \AA~resolution}.
{\it Nature} 389:251-260.

\bibitem{richmond2003}
Richmond TJ, Davey CA (2003)
{The structure of DNA in the nucleosome core}.
{\it Nature} 423:145-150.

\bibitem{white2001}
White CL, Suto RK, Luger K (2001)
{Structure of the yeast nucleosome core particle reveals fundamental changes in internucleosome interactions}.
{\it The EMBO Journal} 20:5207-5218.

\bibitem{rando2010}
Rando OJ (2010)
{Genome-wide mapping of nucleosomes in yeast}.
{\it Methods in Enzymology} 470:105-118.

\bibitem{zinchenko2006}
Zinchenko AA, Chen N (2006)
{Compaction of DNA on nanoscale three-dimensional templates}.
{\it J. Phys.: Condens. Matter} 18:R453-R480.

\bibitem{west2010}
West SM, Rhos R, Mann RS, Honig B (2010)
{Electrostatic interactions between arginines and minor groove in the nucleosome}.
{\it Journal of biomolecular structure \& dynamics} 27:861-866.

\bibitem{fenley2010}
Fenley AT, Adams DA, Onufriev AV (2010)
{Charge state of the globular histone core controls stability of the nucleosome}.
{\it Biophysical Journal} 99:1577-1585.

\bibitem{widom1986}
Widom J (1986)
{Physicochemical studies of the folding of the 100 \AA~nucleosome filament into the 300 \AA~filament}.
{\it J. Mol. Biol.} 190:411-424.

\bibitem{horowitz1994}
Horowitz RA, Agard DA, Sedat JW, Woodcock CL (1994)
{The three-dimensional architecture of chromatin. In situ: Electron tomography reveals fibers composed of a continuously variable zig-zag nucleosomal ribbon}.
{\it The Journal of Cell Biology} 125:1-10.

\bibitem{bednar1998}
Bednar J, Horowitz RA, Grigoryev SA, Carruthers LM, Hansen JC, Koster AJ, Woodcock CL (1998)
{Nucleosomes, linker DNA, and linker histone form a unique structural motif that directs the higher-order folding and compaction of chromatin}.
{\it Proc. Natl. Acad. Sci. USA}  95:14173-14178.

\bibitem{wyrick1999}
Wyrick JJ, Holstege FCP, Jennings EG, Causton HC, Shore D, Grunstein M, Lander ES, Young RA (1999)
{Chromosomal landscape of nucleosome-dependent gene expression and silencing in yeast}.
{\it Nature} 402:418-421.

\bibitem{berger2007}
Berger SL (2007)
{The complex language of chromatin regulation during transcription}.
{\it Nature} 447:407-412.

\bibitem{shukla2009}
Shukla A, Chaurasia P, Bhaumik SR (2009)
{Histone methylation and ubiquitination with their cross-talk and roles in gene expression and stability}.
{\it Cell. Mol. Life Sci.} 66:1419-1433.

\bibitem{ausio1986}
Ausio J, van Holde KE (1986)
{Histone hyperacetylation: its effects on nucleosome conformation and stability}.
{\it Biochemistry} 25:1421-1428.

\bibitem{strahl2000}
Strahl BD, Allis CD (2000)
{The language of covalent histone modifications}.
{\it Nature} 403:41-45.

\bibitem{turner2000}
Turner BM (2000)
{Histone acetylation and an epigenetic code}.
{\it Bioessays} 22:836-845.

\bibitem{sekinger2005}
Sekinger EA, Moqtaderi Z (2005)
{Intrinsic histone-DNA interactions and low nucleosome density are important for preferential accessibility of promoter regions in yeast}.
{\it Molecular cell} 18:735-748.

\bibitem{shilatifard2006}
Shilatifard A (2006)
{Chromatin modiÞcations by methylation and ubiquitination: Implications in the regulation of gene expression}.
{\it Annu. Rev. Biochem.} 75:243-269.

\bibitem{strauss1984}
Strauss F, Varshavsky A (1984)
{A protein binds to a satellite DNA repeat at three specific sites that would be brought into mutual proximity by DNA folding in the nucleosome}.
{\it Cell} 37:889-901.

\bibitem{schiessel2001}
Schiessel H, Widom J, Bruinsma RF, Gelbart WM (2001)
{Polymer reptation and nucleosome repositioning}.
{\it Phys. Rev. Lett.} 86:4414-4417 and Erratum: {\it Phys. Rev. Lett.} 88:129902-1 (2002).

\bibitem{zlatanova2003}
Zlatanova J, Leuba SH (2003)
{Chromatin Fibers, One-at-a-time}.
{\it J. Mol. Biol.} 331:1-19.

\bibitem{schwabish2006}
Schwabish MA, Struhl K (2006)
{Asf1 Mediates histone eviction and deposition during elongation by RNA polymerase II}.
{\it Molecular Cell} 22:415-422.

\bibitem{levitt1978}
Levitt M (1978)
{How many base-pairs per turn does DNA have in solution and in chromatin? Some theoretical calculations}
{\it Proc. Natl. Acad. Sci. USA} 75:640-644.

\bibitem{klug1981}
Klug A, Lutter LC (1981)
{The helical periodicity of DNA on the nucleosome}.
{\it Nucleic Acids Res.} 9:4267-4283.

\bibitem{hayes1990}
Hayes JJ, Tullius TD, Wolffe AP (1990)
{The structure of DNA in a nucleosome}.
{\it Proc. Natl. Acad. Sci. U.S.A.} 87:7405-7409.

\bibitem{white1969} White JH (1969)
{Self-linking and the Gauss integral in higher dimensions}.
Am. J. Math., 91:693-728. 

\bibitem{skjeltorp1996}
Skjeltorp AS, Clausen S, Helgesen G, Pieranski P (1996)
{Knots and applications to biology, chemistry and physics}.
in {\it Physics of biomaterials:fluctuations, selfassembly, and evolution}
eds Riste T, Sherrington DC
{\it NATO ASI Series E} 322 pp 187-217.
(Kluwer Academic Publishers, Dordrecht)


\bibitem{besker2005}
Be\v{s}ker N, Anselmi C, Santis PD (2005)
{Theoretical models of possible compact nucleosome structures}.
{\it Biophysical Chemistry} 115:139-143.

\bibitem{wong2007}
Wong H, Victor J-M, Mozziconacci J (2007)
{An all-atom model of the chromatin fiber containing linker histones reveals a versatile structure tuned by the nucleosomal repeat length}.
{\it PLoS One} 2:e877:1-8.

\bibitem{kepper2008}
Kepper K, Foethke D, Stehr R, Wedemann G, Rippe K (2008)
{Nucleosome geometry and internucleosomal interactions control the chromatin fiber conformation}.
{\it Biophysical Journal} 95:3692-3705.

\bibitem{olsen2009}
Olsen K, Bohr J (2010)
{The generic geometry of helices and their close-packed structures}.
{\it Theor. Chem. Acc.} 125:207-215.

\bibitem{olsenZT}
Olsen K, Bohr J (2011)
{The geometrical origin of the strain-twist coupling in double helices}.
Submitted: For a preprint see http://arxiv.org/abs/1003.5358.

\bibitem{Bohrcollagen}
Bohr J, Olsen K (2011)
{The close-packed triple helix as a possible new structural motif for collagen}.
{\it Theor. Chem. Acc.} DOI: 10.1007/s00214-010-0761-3.

\bibitem{vasudevan2010}
Vasudevan D, Chua EYD, Davey CA (2010)
{Crystal structures of nucleosome core particles containing the 601 strong positioning sequence}.
{\it J. Mol. Biol.} 403:1-10.

\bibitem{pieranski1998}
Piera\'{n}ski P, Przyby\l~S (1998)
{In search of ideal knots}
in {\it Ideal Knots}
eds Stasiak A, Katritch V, Kauffman LH
(World Scientific, Singapore)
ISBN 981-02-3530-5 pp 22-41.

\bibitem{przybyl2001}
Przyby\l~ S, Piera\'{n}ski P (2001)
{Helical close packings of ideal ropes}.
{\it Eur. Phys. J. E} 4:445-449.

\bibitem{neukirch2002}
Neukirch S, van der Heijden GHM (2002)
{Geometry and mechanics of uniform $n$-plies: from engineering ropes to biological filaments}.
{\it Journal of Elasticity}  69:41-72.

\bibitem{lavelle2010}
Lavelle C, Victor J-M, Zlatanova J (2010)
{Chromatin fiber dynamics under tension and torsion}.
{\it Int. J. Mol. Sci.} 11:1557-1579.


\end{thebibliography}
\end{document}